\documentclass[conference,twocolumn]{IEEEtran}
\IEEEoverridecommandlockouts
\usepackage{cite}
\usepackage{amsmath,amssymb,amsfonts}
\usepackage{graphicx}
\usepackage{comment}
\usepackage{textcomp}
\usepackage{siunitx}
\usepackage{xcolor}
\usepackage{tikz}
\usepackage{mathtools}
\usepackage{amsmath, amssymb, cases}  
\usepackage{lipsum}  
\usepackage{gensymb}
\usepackage{wrapfig}
\usepackage{mathrsfs}
\usepackage[caption=false,font=footnotesize]{subfig}
\usepackage{wrapfig}
\usepackage{breqn}
\usepackage{amsfonts}
\usepackage{graphicx, color}
\usepackage[margin=0.5in]{geometry} 

\usepackage{algorithm}
\usepackage[noend]{algorithmic}

\newcommand{\mbf}[1]{\mathbf{#1}}

\def\mbfibTeX{{\rm B\kern-.05em{\sc i\kern-.025em b}\kern-.08em
    T\kern-.1667em\lower.7ex\hbox{E}\kern-.125emX}}
\begin{document}

\newcommand\copyrighttext{%
  \scriptsize \textcopyright 2020 IEEE. Personal use of this material is permitted.  Permission from IEEE must be obtained for all other uses, in any current or future media, including reprinting/republishing this material for advertising or promotional purposes, creating new collective works, for resale or redistribution to servers or lists, or reuse of any copyrighted component of this work in other works.}
\newcommand\copyrightnotice{%
\begin{tikzpicture}[remember picture,overlay]
\node[anchor=south,yshift=1pt] at (current page.south) {\fbox{\parbox{\dimexpr\textwidth-\fboxsep-\fboxrule\relax}{\copyrighttext}}};
\end{tikzpicture}%
}

\author{\IEEEauthorblockN{Gabriel Schamberg$^{1,*}$, Sourish Chakravarty$^{1,*}$, Taylor E. Baum$^2$, Emery N. Brown$^{1,3,4}$}
\IEEEauthorblockA{\\
$^1$\textit{Picower Institute for Learning and Memory}, \textit{Massachusetts Institute of Technology} \\
$^2$\textit{Electrical Engineering and Computer Science Department}, \textit{Massachusetts Institute of Technology} \\
$^3$\textit{Department of Anesthesia, Critical Care and Pain Medicine}, \textit{Massachusetts General Hospital} \\
$^4$\textit{Division of Sleep Medicine}, \textit{Harvard Medical School}\\
MA, USA}
\thanks{$^*$These authors contributed equally to this work.

This work was supported by a Picower Postdoctoral Fellowship (to GS), the National Science Foundation GRFP 1122374 (to TEB), the National Institutes of Health P01 GM118629 (to ENB), and funds from the Picower Foundation (to ENB).}
}

\allowdisplaybreaks



\title{Inferring neural dynamics during burst suppression using a neurophysiology-inspired switching state-space model}

\maketitle
\copyrightnotice
\begin{abstract}
Burst suppression is an electroencephalography (EEG) pattern associated with profoundly inactivated brain states characterized by cerebral metabolic depression. Its distinctive feature is alternation between short temporal segments of near-isoelectric inactivity (suppressions) and relatively high-voltage activity (bursts). Prior modeling studies suggest that burst-suppression EEG is a manifestation of two alternating brain states associated with consumption (during a burst) and production (during a suppression) of adenosine triphosphate (ATP). This finding motivates us to infer latent states characterizing alternating brain states and underlying ATP kinetics from instantaneous power of multichannel EEG using a switching state-space model. Our model assumes Gaussian distributed data as a broadcast network manifestation of one of two global brain states. The two brain states are allowed to stochastically alternate with transition probabilities that depend on the instantaneous ATP level, which evolves according to first-order kinetics. The rate constants governing the ATP kinetics are allowed to vary as first-order autoregressive processes. Our latent state estimates are determined from data using a sequential Monte Carlo algorithm. Our neurophysiology-informed model not only provides unsupervised segmentation of multi-channel burst-suppression EEG but can also generate additional insights on the level of brain inactivation during anesthesia.
\end{abstract}


\section{Introduction}

Burst suppression is an electroencephalography (EEG) pattern comprising alternating periods of low-amplitude (suppressions) and high-amplitude (bursts) voltage activities \cite{clark1973neurophysiologic}. This unique neurophysiological phenomena manifests during profound states of unconsciousness in general anesthesia, medically-induced coma, coma due to diffuse anoxic damage, epilepsy due to Ohtahara syndrome, and induced hypothermia  \cite{brown2010general}. The increase in the length of suppression segments during the burst suppression state is associated with further reduction in neuronal activities and in cerebral metabolism \cite{steriade1994cortical, michenfelder1991relationship}.  Therefore, the fraction of time spent in the suppression epoch is used as a biomarker of level of unconsciousness in profoundly unconscious patients \cite{Rampil1988, chemali_burst_2013}. Realtime monitoring of this biomarker can inform anesthesiologists or closed-loop anesthesia delivery systems to precisely titrate the anesthetic drug dosage so as to maintain a desired level of unconsciousness \cite{Ching2013Real-timeSuppression, shanechi2013brain}.

Burst suppression EEG is often segmented into bursting and suppression epochs by comparing either the maximum voltage in a sliding window of EEG data or a recursive estimate of the local EEG variance to a fixed threshold \cite{brandon_westover_real-time_2013}. A traditional approach to quantify the fraction of the time spent in suppression is via the burst suppression rate (BSR)\cite{Rampil1988}, which calculates the fraction of time that EEG is suppressed over a fixed time window. An alternative to BSR is the burst suppression probability (BSP), which gives the instantaneous probability of suppression, and is calculated by fitting a point process state-space model (SSM) to the binary data segmentation without requiring a subjectively-defined sliding window \cite{chemali_burst_2013}. The BSP metric has a similar interpretation as the BSR, but the former has additional advantages in terms of higher temporal resolution, principled interpretation as a probability measure, robustness to noise in the data, and availability of the joint probability distribution on the entire BSP trajectory, which can be useful for inference. Although the BSP estimate is statistically principled and clinically relevant, it is not designed to provide any neurophysiologic insights.

The multiple etiologies (e.g. in GA, hypothermia, medically-induced coma) of burst suppression  share a common property - all are associated with lowered brain metabolism \cite{ching2012neurophysiological}. Based on this property, Ching et al. posited a neurophysiologic model comprising rhythm-generating neuronal circuits which associates the alternating suppression and burst epochs with regeneration and depletion of adenosine triphosphate (ATP), respectively. Based on this model, a decrease in the rate of ATP production (say, due to increasing doses of anesthetic drug) will result in decreased neuronal spiking activity and longer suppression epochs in the concurrently occurring local field potentials. Prior works have incorporated this neurophysiology-based intuition to develop statistical models that represent the transition dynamics among burst and suppression epochs using two-state semi-Markov process wherein state transitions are governed by the current ATP level. These neurophysiology-informed SSMs allow for the dynamics in the metabolic level, quantified by the ATP production rate, from single-channel burst suppression EEG \cite{Westover2015Real-timeSuppression, chakravarty_hidden_2019}. The advantage of such neurophysiology-informed models is that they can provide mechanistic interpretability of biomarkers estimated from EEG data, and can potentially provide numerical constraints to build patient-specific mechanistic models. However, there are a some key limitations in these SSM-based works. Both \cite{Westover2015Real-timeSuppression, chakravarty_hidden_2019} use segmented sequence as the observation thus requiring a threshold to be prescribed. Furthermore, \cite{Westover2015Real-timeSuppression} requires intermediate BSP estimates to calculate the metabolic level whereas \cite{chakravarty_hidden_2019} can directly estimate the metabolic level from the binarized sequence but assumes fixed burst suppression depths (i.e. ATP production rate) within subjectively prescribed sliding time windows.


Our current work builds on the recent neurophysiology-informed SSM in \cite{chakravarty_hidden_2019} by addressing some of its aforementioned limitations and extending the framework to directly estimate metabolic level dynamics from multi-channel EEG without the need for intermediate segmentation. Our switching SSM assumes a hierarchy of three latent state processes that describe the dynamics in metabolic level, the ATP level, and the discrete state transitions (Section \ref{sec:model}). To estimate the latent state dynamics from a given EEG recording, we have developed particle filtering and smoothing algorithms (Section \ref{sec:particle}). We analyze the properties of the estimation framework using simulated data, demonstrate its utility in analyzing EEG from human subjects under general anesthesia (Section \ref{sec:results}), and finally discuss the implications and next steps of this work (Section \ref{sec:discussion}). 

\section{Model formulation}\label{sec:model}

Let the $k$-th data point in a multi-channel EEG voltage time-series, sampled at $F_s$ (Hz) and detrended, be represented by $\mbf{y}_k\in \mathbb{R}^{N}$, where $N$ denotes the number of channels analyzed. We define the gap between time points $k$ and $(k+1)$ as $\Delta=(1/F_s)$.
Our proposed SSM uses three hidden state processes:
\begin{itemize}
    \item 
$s_k \in \lbrace 1, 2 \rbrace$ represents whether the subject is in a state of bursting ($s_k=1$) or suppression ($s_k=2$).
\item
$x_k\in [0,1)$ indicates a measure of ATP concentration.
\item
$z_k \in \mathbb{R}$ characterizes the log energy production rate, which is inversely related to the depth of the subject's burst-suppression state.
\end{itemize}
Throughout the paper, we use bold letters to refer to vectors, lowercase letters to refer to random variables, and capital letters to refer to samples of those random variables. We use the symbol $p(\cdot)$ to represent both probability density functions for continuous-valued random variables or probability mass functions for discrete-valued random variable, with the specific distributions being made clear by context.
\subsection{Observation model}
For the observations, we assume a multivariate normal distribution with zero mean and a covariance that depends whether the data point belongs to a burst or a suppression segment. We assume conditional independence across the channels given the state:
\begin{align} \label{eq:samples}
    \mbf{y}_k\vert s_k, x_k, z_k = \mbf{y}_k\vert s_k \sim \mathcal{N}(0, \Sigma_{s_k} )
    \\
    \Sigma_{s_k} = diag([\sigma_{s_k,1}^2,\cdots,\sigma_{s_k,N}^2 ]) \nonumber
\end{align}
where $\mbf{y}_k=[y_{k,1},\dots,y_{k,N}]$ and $\sigma_{1,n}^2 \gg \sigma_{2,n}^2$, $\forall\, n=1,\dots,N$. Next, we define the windowed signal power with window length $W$ in channel $n$ as:
\begin{equation}
    \bar{y}_{k',n} = \frac{1}{W} \sum_{w=0}^{W-1} y_{k'-w,n}^2
\end{equation}

\noindent To ensure that the observations $\bar{y}_{k',n}$ are conditionally independent given $s_{1:K}$, we only consider observations at times $k'=W,~2W,\dots,~K$. We further impose the assumption that $s_{k'}=s_{k'-w}$ for $w=1,\dots,W$ (i.e. the down-sampling occurs exactly at state transitions). Under this assumption we have that appropriately scaled versions of the observed powers are conditionally independently distributed according to a chi-squared distribution with $W$ degrees of freedom:
\begin{equation}
    \frac{W}{\sigma_{s_{k'},n}^2}\bar{y}_{k',n} \mid s_{k'} \sim \chi^2_W
\end{equation}
and thus:
\begin{equation}\label{eq:power}
   \bar{y}_{k',n} \mid s_{k'} \sim \Gamma\left(\frac{W}{2},2\frac{\sigma_{s_{k'},n}^2}{W}\right)
\end{equation}
where $\Gamma(\kappa,\theta)$ is the gamma distribution parameterized by shape $\kappa$ and scale $\theta$ factors. When presenting the particle-based inference methods in Section \ref{sec:particle}, we will see that the likelihood models in \eqref{eq:samples} and \eqref{eq:power} can be used interchangeably, requiring only the likelihood evaluation and sampling rates to be adjusted.
\begin{figure}[t]
    \begin{center}
        \includegraphics[width=0.8\linewidth]{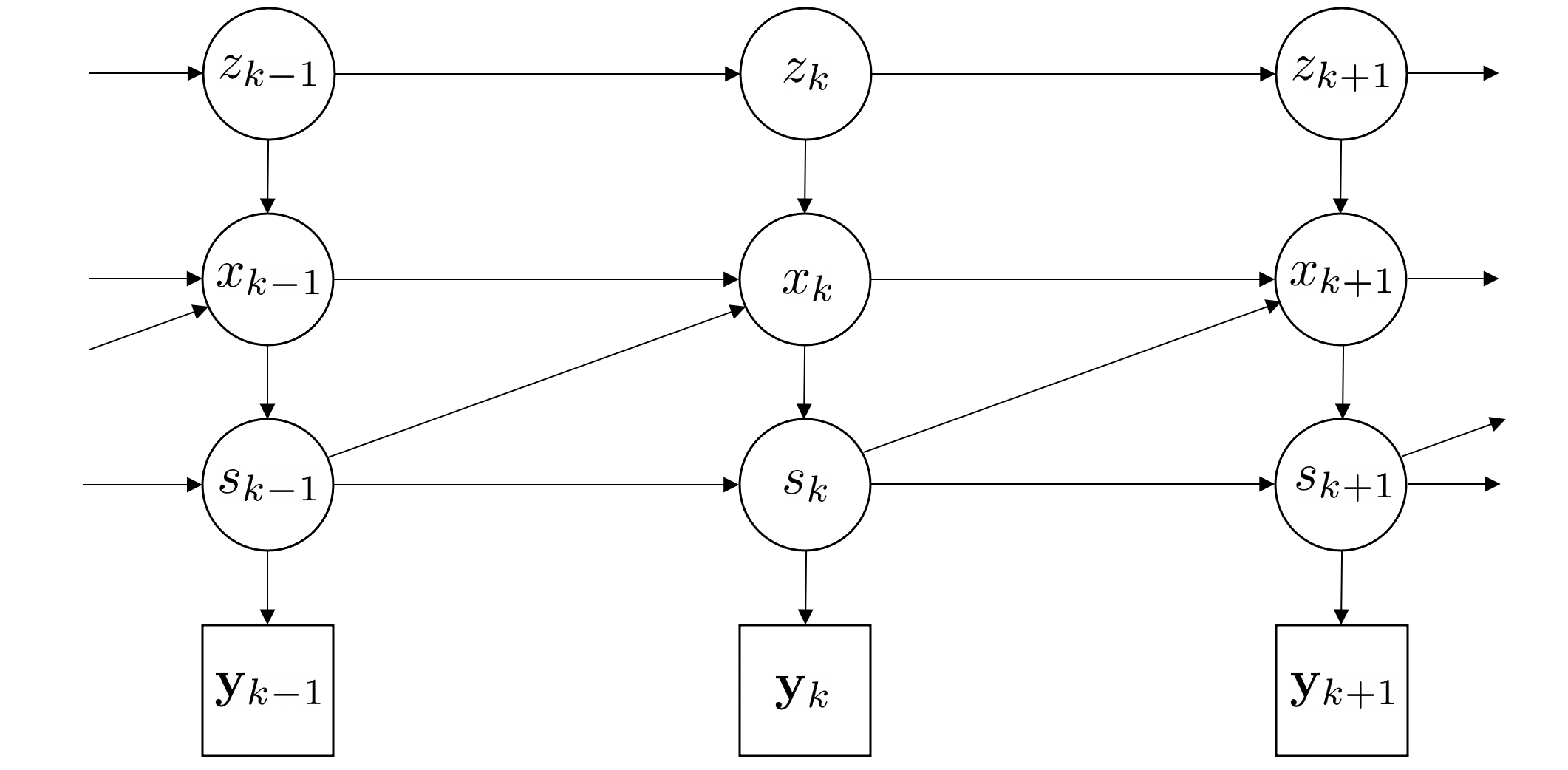}
    \end{center}
    \caption{\label{fig:graphmodel}Probabilistic graphical model representing statistical dependencies among the latent dynamical state and observation processes}
 \end{figure}
 
\subsection{Latent state evolution models}\label{sec:dynamics}
We model the log energy production rate as evolving according to a Gaussian random walk:
\begin{equation}\label{eq:z}
    z_{k} = z_{k-1} + \varepsilon_k^{(z)} 
\end{equation}
\noindent where $\varepsilon_k^{(z)} \sim \mathcal{N}(0, \sigma_{z}^2)$ and $z_{0}\sim \mathcal{N}(\mu_{z_0}, \sigma^2_{z_0})$. 
Motivated by \cite{ching2012neurophysiological} and \cite{westover_real-time_2013}, the normalized energetic state evolves according to a reduced order model where energy is regenerated at all times and consumed only during bursting:
\begin{align}\label{eq:x}
    x_{k} = 
    \begin{cases}
    c_0^1[x_{k-1} + \lambda^{(p)}_k(1-x_{k-1})\Delta - \lambda^{(c)}\Delta  + \varepsilon_k^{(x)}] ,\,\,\, s_{k-1}= 1
    \\
    c_0^1[x_{k-1} + \lambda^{(p)}_{k} (1-x_{k-1}) \Delta + \varepsilon_k^{(x)}] ,\,\,\,\,\,\,s_{k-1}= 2
    \end{cases}
\end{align}
where $\lambda^{(p)}_{k} = \exp( z_{k})$ is non-negative ATP production rate, $\lambda^{(c)}$ denotes a fixed ATP consumption rate, $c_0^1[x] =\min(\max(x,0),1)$ is shorthand for a clipping operation, and $\varepsilon_k^{(x)}\sim\mathcal{N}(0,\sigma^2_x)$. We assume $x_0 \sim Unif[0,1]$. 
Eq.~\eqref{eq:x} indicates that $x_k$ approaches $1$ during suppression segments and $0$ during bursting segments. To account for transitioning between bursting and suppression states, we first define a gating function $g:[0,1]\times \{1,2\}\rightarrow[0,1]$ that maps an energetic level to a transition probability:
\begin{equation}
    g(x_k, s_{k-1})= 
    \begin{cases}
    \frac{x_k^{\gamma_{2}}}{C_{2}^{\gamma_{2}}+x_k^{\gamma_{2}}}, &s_{k-1}=2 
    \\
    1-\frac{x_k^{\gamma_{1}}}{C_{1}^{\gamma_{1}}+x_k^{\gamma_{1}}}, &s_{k-1}=1 
    \end{cases}
     \label{eq:g_k}
\end{equation}
where $C_{s_k}$ denotes the energetic level corresponding to a 50\% chance of transitioning out of state $s_k$ and $\gamma_{s_k}$ determines the steepness of the gating function. 
The gating function is used to determine the switching probability of the discrete process $s_k$:
\begin{equation}\label{eq:s}
    \text{Pr}(s_{k}\ne s_{k-1}\mid s_{k-1},x_{k}) = g(x_{k},s_{k-1})
\end{equation}
where $\text{Pr}(\cdot)$ indicates the relevant probability measure and $\text{Pr}(s_0=1)=\pi_1$.
The statistical dependencies between the three latent processes are illustrated as a graphical model in Fig.~\ref{fig:graphmodel}. Importantly, we see the complete data likelihood factorizes as:
\begin{align}\label{eq:chain_rule}
   &  p(\mbf{y}_{1:K},s_{0:K},x_{0:K},z_{0:K}) =
    p(z_0, x_0, s_0)
    \times
    \\
   &  \prod_{k=1}^K p(z_k \mid z_{k-1}) p(x_k \mid z_k,x_{k-1},s_{k-1}) p(s_k \mid x_k,s_{k-1}) p(\mbf{y}_k \mid s_k) 
   \nonumber
\end{align}
where we assume $p(z_0, x_0, s_0)=p(z_0)p(x_0)p(s_0)$. For brevity, we define a latent state $\mbf{q}_k\triangleq(z_k,x_k,s_k)$ that allows us to rewrite Eq.~\eqref{eq:chain_rule} as $p(\mbf{y}_{1:K},\mbf{q}_{0:K}) =p(\mbf{q}_0)  \nonumber \prod_{k=1}^K p(\mbf{q}_k \mid \mbf{q}_{k-1})p(\mbf{y}_k \mid \mbf{q}_k)$. 
\section{State Estimation}\label{sec:particle}
Given the highly nonlinear dynamics in our proposed model, we adopt a sequential Monte Carlo (SMC) or Particle Filter-based approach to estimate the posterior distribution $p(\mbf{q}_{1:K}\mid \mbf{y}_{1:K})$\cite{doucet_tutorial_2009}\footnote{SMC methods are desirable here because they afford a reasonably straightforward implementation and do not require linearization or functional approximation. The drawback of these methods is computational complexity, which can be prohibitive in cases where many particles are required to obtain an accurate estimate of the posterior.}. In particular, we implement a forward filter/backward smoother approach wherein a forward pass produces a collection of weighted sample trajectories of the latent states, and a backward recursion smooths the weights without drawing any new samples 
The complete set of model parameters for our proposed statistical model, described by Eqs.~\eqref{eq:samples}, \eqref{eq:z}, \eqref{eq:x},  \eqref{eq:g_k} and \eqref{eq:s}, is given by  $\Theta = \lbrace \lbrace \sigma^2_{1,n}\rbrace_{n=1}^N, \lbrace \sigma^2_{2,n}\rbrace_{n=1}^N$, $\mu_{z_0}, \sigma_{z_0}^2, \sigma_{z}^2,\sigma_{x}^2, C_1, \gamma_1, C_2, \gamma_2, \pi_1 \rbrace$. For this complex statistical model  we treat the model parameters as tuning parameters. 
\subsection{Filtering}

\begin{algorithm}[t]
\caption{Particle filter}\label{alg:filter}
 \begin{algorithmic}[1]
\renewcommand{\algorithmicrequire}{\textbf{Input:} $J$, $L$, $\Theta$, and $\mbf{y}_{1:K}$}
\renewcommand{\algorithmicensure}{\textbf{Output:} $\mbf{Q}_{1:K}^j$ and $\bar{\alpha}^j_{1:K}$ for $j=1,\dots, J$}
 \REQUIRE 
 \ENSURE  
  \STATE $Z^j_{0}\sim \mathcal{N}(\mu_{z_0},\sigma^2_{z_0})$
  \COMMENT{draw $z_0$ samples}
  \STATE $X^j_{0}\sim Unif[0,1]$
  \COMMENT{draw $x_0$ samples}
  \STATE $S^j_{0}\sim Bern(\pi_1)$
  \COMMENT{draw $s_0$ samples}
  \STATE $\bar{\alpha}_k^j=\frac{1}{J}$
  \COMMENT{initialize uniform weights}
  \FOR {$k=1,\dots,K$}
  \STATE $Z^j_{k}\sim p(\cdot \mid Z^j_{k-1})$
  \COMMENT{sample using \eqref{eq:z}}
  \STATE $X^j_{k}\sim p(\cdot \mid Z^j_{k},X^j_{k-1},S^j_{k-1})$
  \COMMENT{sample using \eqref{eq:x}}
  \STATE $S^j_{k}\sim p(\cdot \mid X^j_{k},S^j_{k-1})$
  \COMMENT{sample using \eqref{eq:s}}
  \STATE $\alpha^j_{k} = \bar{\alpha}^j_{k-1} p(\mbf{y}_k \mid S^j_{k})$
  \COMMENT{reweight using \eqref{eq:samples} or \eqref{eq:power}}
  \STATE $\bar{\alpha}^j_{k} = \frac{\alpha^j_k}{\sum_{l=1}^J\alpha^l_{k}} $
  \COMMENT{normalize weights}
  \IF {$ESS(\bar{\boldsymbol{\alpha}}_k)<\frac{J}{2}$}
  \STATE $\{\bar{\alpha}_{k-L:k}^j,\mbf{Q}_{k-L:k}^j\} \coloneqq \{\frac{1}{J},\tilde{\mbf{Q}}_{k-L:k}^j\}$
  \COMMENT{resample}
  \ENDIF
  \ENDFOR
 \end{algorithmic} 
 \end{algorithm}

We first consider the problem of obtaining samples from our target density $p(\mbf{q}_{1:k} \mid \mbf{y}_{1:k})$ in an online fashion. The key component of SMC algorithms is a \emph{proposal density} from which we can sequentially draw samples. Here we leverage the latent state evolution models described in Section \ref{sec:dynamics} to set the proposal density as $p(\mbf{q}_{1:K})=p(\mbf{q}_{1})\prod_{k=2}^{K} p(\mbf{q}_{k}\mid \mbf{q}_{1:k-1})$. Thus, supposing at time $k$ we have $J$ samples given by $\{\mbf{Q}_k^j\}_{j=1}^J=\{(Z_k^j,X_k^j,S_k^j)\}_{j=1}^J$, the samples at time $k+1$ are generated as $\mbf{Q}_{k+1}^j\sim p(\cdot \mid \mbf{Q}_{k}^j)$.

To account for the observed data in the posterior state estimation, each sample is allocated a weight that can be derived using standard SMC techniques \cite{doucet_tutorial_2009}. Specifically, we note that the target density can be written as:
\begin{equation}
    p(\mbf{q}_{1:k} \mid \mbf{y}_{1:k}) 
    = \frac{p(\mbf{q}_{1:k},\mbf{y}_{1:k})}{p(\mbf{q}_{1:k})} \frac{p(\mbf{q}_{1:k})}{p(\mbf{y}_{1:k})}
    \triangleq  \alpha_k \frac{p(\mbf{q}_{1:k})}{p(\mbf{y}_{1:k})}
\end{equation}
where $\alpha_k=p(\mbf{y}_{1:k}\mid\mbf{q}_{1:k})$ represents the \emph{unnormalized weight function}. This can be expanded as: 
\begin{align}
   & \alpha_k =
    \frac{p(\mbf{q}_{1:k-1},\mbf{y}_{1:k-1})}{p(\mbf{q}_{1:k-1})}
    \frac{p(\mbf{q}_{1:k},\mbf{y}_{1:k})}{p(\mbf{q}_{1:k-1},\mbf{y}_{1:k-1})p(\mbf{q}_k \mid \mbf{q}_{1:k-1})} 
    \\
    &=
    \alpha_{k-1}
    \frac{p(\mbf{q}_{k},\mbf{y}_{k}\mid \mbf{q}_{1:k-1},\mbf{y}_{1:k-1})}{p(\mbf{q}_k \mid \mbf{q}_{1:k-1})} 
    =
    \alpha_{k-1}
    \frac{p(\mbf{q}_{k},\mbf{y}_{k}\mid \mbf{q}_{1:k-1})}{p(\mbf{q}_k \mid \mbf{q}_{1:k-1})} \label{eq:indp1}\\
    &=
    \alpha_{k-1}
    p(\mbf{y}_{k}\mid \mbf{q}_{1:k}) =
    \alpha_{k-1}
    p(\mbf{y}_{k}\mid s_k) \label{eq:indp2}
\end{align}
\noindent where \eqref{eq:indp1} and \eqref{eq:indp2} follow from the conditional independence relationships indicated by Fig. \ref{fig:graphmodel}. Thus, the weights associated with each sample are computed recursively by multiplying the previous weight by the likelihood of the newly observed EEG. Since both $z_k$ and $x_k$ are conditionally independent of $\mbf{y}_k$ given $s_k$, they are only reflected in the weighting insofar as ``good'' samples $Z^j_k$ and $X^j_k$ are more likely to generate a highly weighted sample $S^j_k$. For the particle filter, each sample $\mathbf{Q}_k^j$ is assigned a weight $\alpha^j_k$. Once the weights of all samples at time $k$ are computed, they are normalized to yield $\bar{\alpha}^j_k=\alpha_k^j(\sum_{l=1}^J \alpha^l_k)^{-1}$.

At each time $k$ we evaluate an \emph{effective sample size} of the normalized weights, given by $ESS(\bar{\boldsymbol{\alpha}}_k) = (\left|\left|\bar{\boldsymbol{\alpha}}_k\right|\right|_2^2)^{-1}$, where $\bar{\boldsymbol{\alpha}}_k=[\bar{\alpha}_k^1,\dots,\bar{\alpha}_k^J]$ is the vector of normalized weights. The $ESS$ ranges between $1$ to $J$ where $ESS=1$ indicates that only a single sample has non-zero weight and $ESS=J$ indicates that all samples are equally weighted. Thus $ESS$ provides a sense of the number of samples that represent the data well. When the $ESS$ is below $\frac{J}{2}$ we use {\it systematic resampling} to eliminate samples that poorly represent the observations and duplicate those that represent it well \cite{kitagawa1996monte}. We refer to the collection of samples produced by a round of resampling as $\{\tilde{\mbf{Q}}_k^j\}_{j=1}^J$. To avoid sample degeneracy in the early portions of the sample trajectories resulting from frequent resampling, we used a fixed-lag approximation where only recent samples and weights are updated if resampling is performed \cite{kitagawa2001monte}. The forward filtering algorithm is provided in detail in Algorithm \ref{alg:filter}. Upon completing the filtering step, we can summarize the target distribution using estimates of the weighted average and weighted percentiles.

\subsection{Smoothing}

\begin{algorithm}[t]
\caption{Particle smoother}\label{alg:smoother}
  \begin{algorithmic}[1]
\renewcommand{\algorithmicrequire}{\textbf{Input:} $\mbf{Q}_{1:K}^j$ and $\bar{\alpha}^j_{1:K}$ for $j=1,\dots, J$}
\renewcommand{\algorithmicensure}{\textbf{Output:} $\bar{\alpha}^j_{k\mid K}$, for $j=1,\dots,J$ and $k=1,\dots,K$}
 \REQUIRE 
 \ENSURE 
  \STATE $\bar{\alpha}^j_{K\mid K} = \bar{\alpha}^j_{K}$ for $j=1,\dots,J$
  \COMMENT {initialize final weights}
  \FOR {$k=K-1,\dots,1$}
  \FOR {$l=1,\dots,J$}
  \STATE $\eta^l_k = \sum_{m=1}^J \bar{\alpha}_k^m p(\mbf{Q}^l_{k+1}\mid \mbf{Q}^m_k)$
  \COMMENT {approximate integral in Eq. \eqref{eq:integral}}
  \ENDFOR
  \FOR {$j=1,\dots,J$}
  \STATE $\bar{\alpha}^j_{k\mid K} = \bar{\alpha}_k^j \sum_{l=1}^J \frac{\bar{\alpha}_{k+1\mid K}^l p(\mbf{Q}^l_{k+1}\mid \mbf{Q}^j_k)}{\eta^l_k}$
  \COMMENT {smooth weights}
  \ENDFOR
  \ENDFOR
 \end{algorithmic} 
 \end{algorithm}

Once an entire time course of observations has been collected, it may be desirable to update all weights or samples in a post-hoc manner. Here we use a weight smoothing approach which uses the stored samples from the filtering algorithm to update the weights in a backwards recursion \cite{hurzeler1998monte}. Specifically, note that: 
\begin{align}
    p(\mbf{q}_{1:K} \mid \mbf{y}_{1:K})
    &=p(\mbf{q}_K \mid \mbf{y}_{1:K}) \prod_{k=K-1}^1 p(\mbf{q}_k \mid \mbf{q}_{k+1:K} , \mbf{y}_{1:K})\nonumber \\
    &=p(\mbf{q}_K \mid \mbf{y}_{1:K}) \prod_{k=K-1}^1 p(\mbf{q}_k \mid \mbf{q}_{k+1} , \mbf{y}_{1:k}) \label{eq:indp3}
\end{align}
where \eqref{eq:indp3} follows from the conditional independence relationships indicated by Fig. \ref{fig:graphmodel}. Using Bayes' rule, we have:
\begin{align}
    p(\mbf{q}_k \mid \mbf{q}_{k+1} , \mbf{y}_{1:k})
    &=\frac{p(\mbf{q}_{k+1} \mid \mbf{q}_k,\mbf{y}_{1:k})p(\mbf{q}_{k} \mid \mbf{y}_{1:k})}{p(\mbf{q}_{k+1} \mid \mbf{y}_{1:k})} \label{eq:smoothweight}\\
    &=\frac{p(\mbf{q}_{k+1} \mid \mbf{q}_k)p(\mbf{q}_{k} \mid \mbf{y}_{1:k})}{\int p(\mbf{q}_{k+1} \mid \mbf{q}_k)p(\mbf{q}_{k} \mid \mbf{y}_{1:k}) d\mbf{q}_k}\label{eq:integral}.
\end{align}
We define the quantity on the left hand side of \eqref{eq:smoothweight} to be the smoothed weight $\alpha_{k\mid K}$. We can use this to obtain a smoothed weight $\bar{\alpha}_{k\mid K}^j$ for a sample $\mbf{Q}_k^j$ by evaluating \eqref{eq:smoothweight} for each sample $\mbf{Q}_{k+1}^l$ and taking a weighted sum using $\bar{\alpha}_{k+1\mid K}^l$ for $l=1,\dots,j$.
These steps are provided in detail in Algorithm \ref{alg:smoother}. This computational complexity of this algorithm is $\mathcal{O}(J^2 K)$, thus using only the forward pass fixed lag approximation may be desirable for prohibitively long recordings.

\section{Results\label{sec:results}}
We tested our approach on both simulated and real datasets. In both cases, we used the windowed EEG power likelihood given by \eqref{eq:power} to weight samples. 
\subsection{Simulations}
Using the probabilistic model described in Section \ref{sec:model}, we generated 400 seconds of latent state trajectories and three channels of simulated EEG with the following parameters: $ \sigma^2_{1,n}=445$, $\sigma^2_{2,n}=125$ (for $n=1,2,3$), $\mu_{z_0}=-2$, $\sigma_{z_0}^2=10^{-5}$, $\sigma_{z}^2=10^{-5}$, $\sigma_{x}^2=10^{-5}$,   $C_1=0.01$, $\gamma_1=15$, $C_2=0.99$, $\gamma_2=15$, and $\pi_1=0.5$.  While the probabilistic model is not able to generate realistic EEG traces since the generated data lacks oscillatory components, the simulations allows us to test whether the particle smoother can accurately recover the dynamics in the ATP level and log production rate despite their conditional independence from the observation given the binary segmentation.

\begin{figure}[t]
    \begin{center}
        \includegraphics[width=\linewidth]{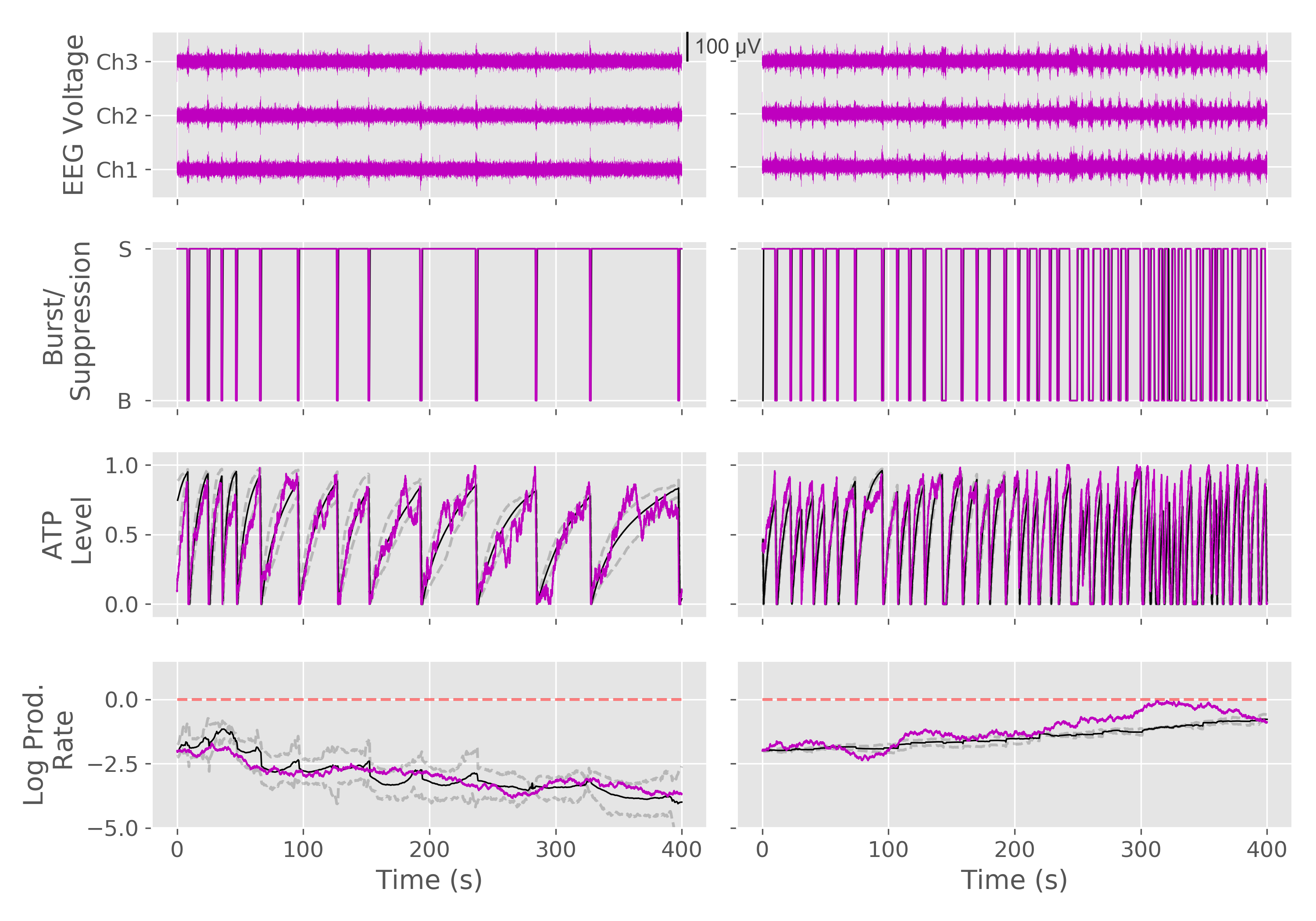}
    \end{center}
    \caption{\label{fig:simulations}Latent state estimation on two realizations of simulated data corresponding to deep burst suppression (left) and light burst suppression (right). The top row shows three channels of observed EEG, with each offset by 200 $\micro$V. The bottom three rows show the true latent states (magenta) and the smoothed estimates (black) with weighted 5th and 95th percentiles (grey). The dashed red line in the bottom row shows the constant ATP log consumption rate.}
\end{figure}

Figure \ref{fig:simulations} shows the generated data, true latent processes, and smoothed estimates for a one simulation exhibiting deep burst suppression and one showing shallow burst suppression. For the deep burst suppression case (Figure \ref{fig:simulations}, left column), we performed the filtering/smoothing with a $\sigma_z^2$ value 10$\times$ \emph{larger} than that used in the data generating model, and likewise 10$\times$ \emph{smaller} for the shallow burst suppression case (Figure \ref{fig:simulations}, right column). Unsurprisingly, the estimated segmentation and ATP levels match the ground truth very closely. The estimated log production rate also tracks the truth reasonably well, although effect of mismatched parameters is apparent in the under- and over-smoothing in the deep and shallow burst suppression simulations, respectively.
\subsection{Human Subjects}
We tested the estimation framework on human EEG data collected from healthy volunteers undergoing propofol-induced anesthesia \cite{purdon_electroencephalogram_2013}. For the observations we used EEG from three frontal electrodes high-pass filtered above 5 Hz. The estimation was performed using the following parameters: $\sigma^2_{1,n}=200$, $\sigma^2_{2,n}=50$ (for $n=1,2,3$), $\mu_{z_0}=0$, $\sigma_{z_0}^2=0.1$, $\sigma_{z}^2=10^{-7}$, $\sigma_{x}^2=10^{-7}$,   $C_1=0.05$, $\gamma_1=13$, $C_2=0.95$, $\gamma_2=13$, and $\pi_1=0.5$. We selected a 2000 second segment of EEG recording that contained varying depths of burst suppression. Given the long duration of this segment, we opted to use only the filtering algorithm in estimating the latent states.

\begin{figure}[t]
    \begin{center}
        \includegraphics[width=\linewidth]{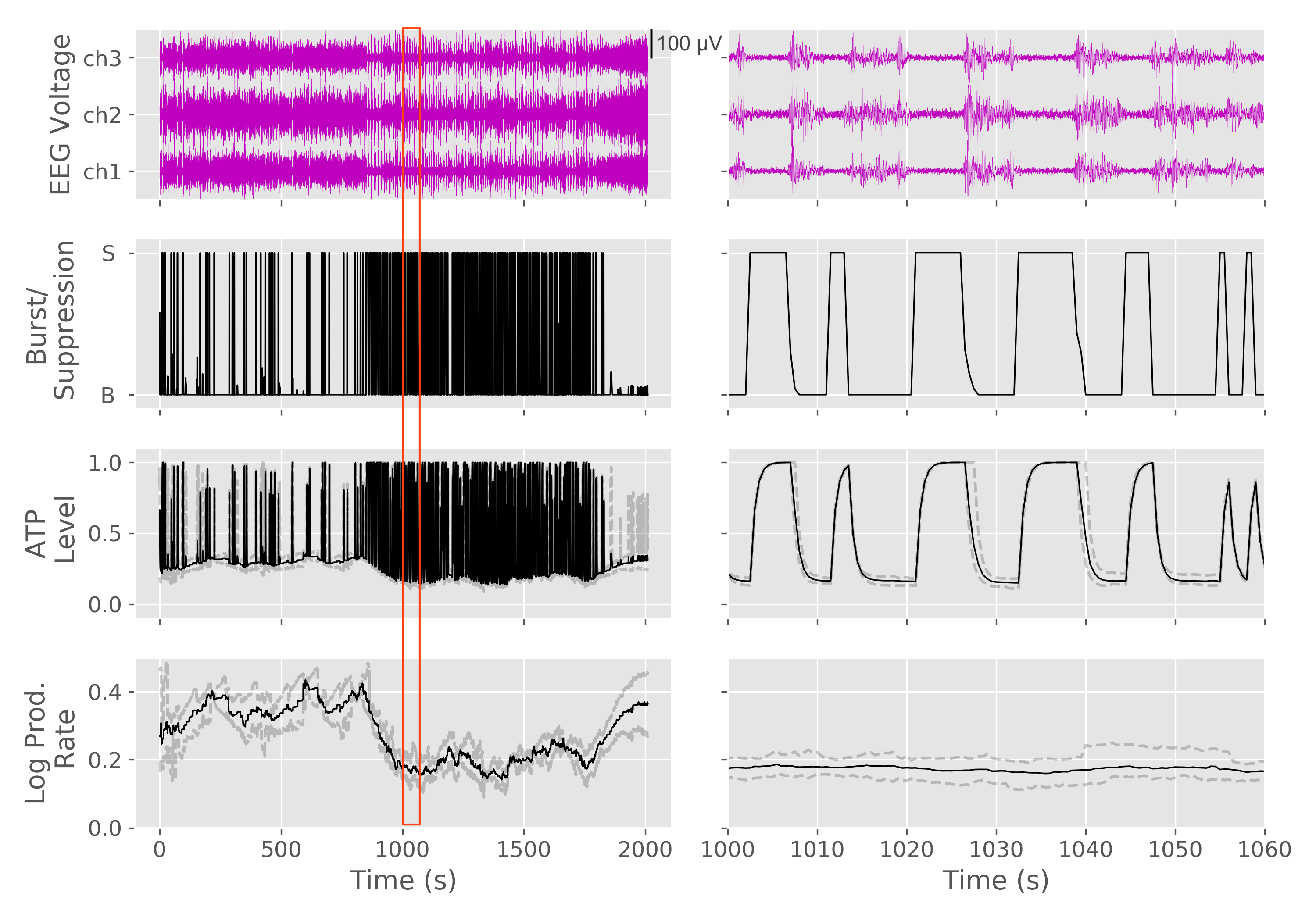}
    \end{center}
    \caption{\label{fig:humans}Latent state estimation on a segment of human burst suppression EEG. The top row displays filtered EEG from the three channels, with each offset by 200 $\micro$V. The bottom three rows show the particle filter latent state estimates with weighted 5th and 95th percentiles. The right column is an enlargement of the red box in the left column.}
\end{figure}

Figure \ref{fig:humans} shows the results of our estimation on the whole data segment and a zoomed in portion of relatively deep burst suppression. We can see that when the subject begins producing suppression epochs more regularly (around 800 seconds), the estimated ATP production rate drops accordingly. In the zoomed panels, we can see that the estimated segmentation tracks nicely with the clearly visible bursting and suppression epochs. This suggests that the estimated ATP production rate can serve as a neurophysiologically-grounded analogue of the abstract BSR/BSP measures. In the shallow burst suppression stage at the beginning of the recording, we can see that the estimated segmentation was somewhere in between burst and suppression, indicating that the samples were not heavily weighted to one of the two states. While this does not happen very frequently, it demonstrates a benefit of the proposed algorithm, namely that uncertainty in what should or should not be deemed a suppression is incorporated into the associated estimation of the other latent states. This is not the case in algorithms that handle segmentation and tracking of burst suppression state independently of one another.

\section{Discussion\label{sec:discussion}}

There are a number of opportunities for continued improvement of the proposed approach. From a modeling perspective, we have made the simplifying assumption that burst suppression is a global brain state. It is known that this is not the case in general, as burst-suppression patterns can occur asynchronously across the cortex \cite{lewis2013local}. As such, future iterations of the model can benefit from using different likelihood functions to capture the spatial properties of burst suppression. Further modeling improvements can be made to the ATP dynamics, which are currently not designed to account for sustained high-amplitude activity (i.e. non-burst-suppression unconsciousness). Additionally, in the context of drug-induced burst suppression, it may be beneficial to augment the model to incorporate drug infusion history in the equation determining the evolution of the ATP production rate using pharmacokinetic/pharmacodynamic modeling as in \cite{yang2016adaptive}.

To improve the estimation framework, it will be important to develop rigorous methods for identifying model parameters. Furthermore, the proposed smoothing algorithm is both computationally demanding and does not resample, requiring that the samples generated in the filtering step have sufficient coverage of the latent state-space. As such, further developments include implementing smoothing approaches that resample the data in a computationally efficient manner \cite{klaas_fast_2006}.

When testing the estimation framework with various parameterizations on the human recordings, we found that, while the estimated ATP level and production rate would change considerably, there was very little deviation in the estimated segmentation. This means that our current paradigm allows for multiple latent trajectories to produce the same burst-suppression pattern. While the proposed method successfully segments the bursts and suppressions, and captures some meaningful dynamics in the depth of the burst suppression, it is clear that more rigorous constraints on what constitutes viable metabolic dynamics need to be established. An important step towards that direction will be to conduct experimental studies where EEG is simultaneously recorded alongside a measure of metabolic activity. Data from such experiments can be useful to validate that the latent processes being estimated by our algorithm do in fact reflect the physiological processes that produce burst-suppression patterns.


\bibliographystyle{ieeetr}
\bibliography{ref}
\end{document}